\documentstyle[11pt,amsmath,amssymb,epsf,epic,cite]{article}
\numberwithin{equation}{section}
\newcommand{\rr}{{\vec R}}
\renewcommand{\ss}{{\vec S}}

\newcommand{\be}{\begin{equation}}
\newcommand{\ee}{\end{equation}}
\newcommand{\bea}{\begin{eqnarray}}
\newcommand{\eea}{\end{eqnarray}}

\newcommand{\cg}{{\ \cal G}\!}

\renewcommand{\c}{{\rm c}}
\newcommand{\e}{{\rm e}}

\renewcommand{\d}{{\rm d}}

\newcommand{\av}[1]{\left\langle{#1}\right\rangle}

\newcommand{\usigma}{{\underline{\sigma \!}}}
\newcommand{\utau}{{\underline{\tau \!}}}

\newcommand{\urho}{{\underline\varrho}}

\newcommand{\grintl}{[\kern-.18em [}
\newcommand{\grintr}{]\kern-.18em ]}

\newcounter{resultcounter}[section]

\newtheorem{thm}[resultcounter]{Theorem}

\newtheorem{prop}[resultcounter]{Proposition}

\newtheorem{definition}[resultcounter]{Definition}

\def\bed{\begin{definition}}
\def\eed{\end{definition}}

\newcommand{\r}{{\rm R}}
\newcommand{\s}{{\rm S}}

\newcommand{\cx}{{\mathbb C}}
\newcommand{\rx}{{\mathbb R}}
\renewcommand{\i}{{\rm i}}

\newcommand{\fer}[1]{(\ref{#1})}
\newcommand{\scalprod}[2]{\left\langle {#1}, {#2}\right\rangle}
\newcommand{\bbbone}{\mathchoice {\rm 1\mskip-4mu l} {\rm 1\mskip-4mu l}
{\rm 1\mskip-4.5mu l} {\rm 1\mskip-5mu l}}

\newcommand{\hhbar}{{}}

\begin{document}

\title{Application of Resonance Perturbation Theory to Dynamics of Magnetization in
Spin Systems Interacting with Local and Collective Bosonic Reservoirs}

\author{M. Merkli\footnote{Email: merkli@mun.ca, http://www.math.mun.ca/$\sim$merkli/ } \\
{\small Dept. of Mathematics and Statistics}\\ {\small  Memorial
University of Newfoundland} \\ {\small St. John's, NL, Canada A1C
5S7} \and  G.P.
Berman\footnote{Email: gpb@lanl.gov}\ \  and A. Redondo\footnote{Email: redondo@lanl.gov}\\
{\small Theoretical Division, MS B213}\\
{\small  Los Alamos National
Laboratory}\\
{\small  Los Alamos, NM 87545, USA}}
\date{\today}
\maketitle
\vspace{-.5cm}

 \markright{right_head}{LA-UR 11-00280}

\begin{abstract}
We apply our recently developed resonance perturbation theory to
describe  the dynamics of magnetization in paramagnetic spin systems
interacting simultaneously with local and collective bosonic
environments. We derive explicit expressions for the evolution of
the reduced density matrix elements. This allows us to calculate
explicitly the dynamics of the macroscopic magnetization, including
characteristic relaxation and dephasing time-scales. We demonstrate
that collective effects (i) do not influence the character of the
relaxation processes but merely  renormalize the relaxation times,
and (ii) significantly modify the dephasing times, leading in some
cases to a complicated (time inhomogeneous) dynamics of the
transverse magnetization, governed by an effective time-dependent
magnetic field.
\end{abstract}

\thispagestyle{empty}
\setcounter{page}{1}
\setcounter{section}{1}
\setcounter{section}{0}

\section{Introduction}

When quantum systems interact with their environments the effects of
relaxation and decoherence occur
\cite{zur,leg,zol,weiss,chuang,SMS,Leggett,PSE}.  In this paper we
study relaxation and decoherence  in quantum macroscopic systems of
``effective" spins interacting simultaneously with both local and
collective thermal environments. By ``effective" we mean that our
approach can be applied not only to magnetic spin systems, but also
to many quantum systems with discrete energy levels, including
recently widely discussed quantum bits (qubits) based on
superconducting Josephson junctions and SQUIDs
\cite{MSS,YHCCW,DM,Setal,Ketal,Cletal}. We also would like to
mention here the research on ephaptic coupling of cortical neurons,
when both local and collective electrical fields play a significant
role in the synchronization dynamics of neurons \cite{CAA}.  We
assume that spins do not interact directly among themselves, but
only through their interactions with collective (energy conserving
and energy exchange) bosonic environments (``thermal baths",
``reservoirs"). The relaxation in these systems is caused by energy
exchange between the environments and spins. The rate of relaxation
is usually characterized by the spectral density of noise of the
reservoirs at the transition frequency, $\omega$, of spins in their
effective magnetic field, and by the interaction constant between a
spin and an environment \cite {Leggett,MSS,Cletal}. The rate of
decoherence usually has a more complicated dependence on the
parameters of spins, their local environments, and interaction
constants \cite {Leggett,PSE,MSS,Cletal,MSB1}. In particular,
low-frequency noise ($1/f$ noise) makes a significant contribution
to the decoherence rate \cite{koch,yosh,bial,shnir} (see also
references therein).

Usually relaxation and decoherence are unwanted effects, for
example, in a quantum computer one must maintain quantum coherence
for long times \cite{chuang}. But dissipative effects can also be
put to good use, for example, in magnetic resonance imaging (MRI)
\cite{clarke,espy,chary}. Indeed, in this case, different values of
relaxation times ($T_1$) for different substances (e.g. water  and
biological tissues) allow one to distinguish and visualize
pathological developments in tissues \cite{clarke}. Dissipative
 effects can also be utilized, for example, to analyze and classify the
 influence of many types of  defects and impurities, in order to
improve the properties of materials \cite{pappas}.

Improving our understanding of relaxation and decoherence processes
is important for many fields of science and for many applications.
The main problem associated  with dissipative effects is that there
are many different sources of noise and thermal fluctuations which
lead to relaxation and decoherence. We mention only some of them,
electromagnetic and acoustic fluctuations (bosonic degrees of
freedom), magnetic fluctuations (such as two-level systems in
superconducting materials), charge defects, and non-equilibrium
quasiparticles.  Generally, it is impossible to eliminate all
sources of noise, so some additional classification can be useful.
For example, in \cite{MSB1} we demonstrated that, for a system of
$N$ spins interacting with  bosonic environments, one can introduce
clusters of reduced density matrix elements in such a way that to a
given  cluster corresponds a decoherence rate describing the fading
of all matrix elements belonging to it. When dealing with a quantum
algorithm in quantum computation, this could imply that the decay of
some clusters is rapid, but -- if the algorithm is built mainly on
the use of slower decaying clusters -- that decay may not influence
significantly the fidelity of the quantum protocol.

In this paper we are mainly interested in the effects produced by
simultaneous influence of both local and collective bosonic
environments on the dynamics of a collective magnetization in a
system of $N$ non-interacting (paramagnetic) spins in a
time-independent magnetic field.  The local and collective
environments
   include both energy conserving and
   energy exchange interactions with spins.  This allows us to determine  conditions
   of applicability of the
   Bloch equation for describing the evolution of the magnetization. We also consider two
   (and more)
   ensembles of spins with different parameters and strengths of interactions
   with their  environments.
   Using our approach based on resonant perturbation theory \cite{MSB1},
   we derive explicit expressions for the time evolution of the reduced
   density matrix elements and, consequently, for the macroscopic magnetization.
   We explicitly calculate the relevant relaxation and decoherence
   rates. The obtained results are important for many applications including MRI and for studying
   collective effects in materials for superconducting qubits.

\bigskip

{\bf Main results of the paper}

\bigskip

$\bullet$ {\it Single spin dynamics.\ } We consider a microscopic, Hamiltonian model of $N$ spins interacting with local and collective bosonic thermal reservoirs, via energy conserving and energy exchange interactions. In Theorem \ref{thm1} we derive a rigorous expression for the reduced density matrix of a single spin, consisting of a main term describing relaxation and dephasing, plus a remainder term which is small in the couplings {\it homogeneously} in time.

$\bullet$ {\it Single spin relaxation.\ } We show that the single-spin relaxation rate is given by
\begin{equation*}
\gamma_{{\rm relax}} = \frac14 \coth(\beta \hhbar\omega/2) \left\{ \lambda^2 J_{g_\c}(\hhbar\omega)+\mu^2 J_{g_\ell}(\hhbar\omega)\right\},
\end{equation*}
where $\omega$ is the spin frequency, $\lambda$ and $\mu$ are the strengths of the energy exchange collective and local couplings, respectively, and where $J_g(\omega)$ is the reservoir spectral density. Only energy-exchange couplings contribute to this rate, and the effect of the local and the collective reservoirs are the same.

$\bullet$ {\it Single spin dephasing.\ } We show that the single-spin dephasing rate is given by
\begin{equation*}
\gamma_{{\rm deph}} = \frac{1}{2}\gamma_{{\rm relax}}+\gamma_{{\rm cons}}+\gamma',
\end{equation*}
where $\gamma_{\rm cons}$ is a contribution stemming only from the
energy  conserving local and collective interactions, determined by
the spectral density of the reservoir at zero frequency (see
\fer{mm-5}). The contribution $\gamma'$ encodes the effect on
dephasing of a single spin due to all other spins. It is defined as
follows. The time-dependence of the single spin off-diagonal density
matrix elements has a very complicated, not exponentially decaying
contribution coming from the collective coupling. The term $\gamma'$
is defined to be the reciprocal of the time by which that quantity
is reduced to half its initial value.

The explicit expression of $\gamma'$ is not simple (see
\fer{preratio}, \fer{c'}).  For small ratio $r$ between the
strengths of the collective to the local couplings we have
$\gamma'=O(r^2)$ (independent of the number $N$ of spins). For large
collective coupling we have $\gamma'\sim {\rm const.} \gamma_{\rm
relax}$, for a constant not depending on $N$.

$\bullet$ {\it Evolution of magnetization.\ } We consider the spins
in a homogeneous magnetic  field pointing in the $z$-direction. We
show that the $z$-component of the total magnetization vector
relaxes to its equilibrium value at the single-spin relaxation rate
$\gamma_{\rm relax}$. This verifies the correctness of the usual
Bloch equation \fer{++} for the $z$-component. The equation for the
transverse total magnetic field is given by a modified Bloch
equation \fer{105}, with a time-dependent dephasing time
($T_2=T_2(t)$) and a time-dependent effective magnetic field. For
large times, the coefficients in the modified Bloch equation
approach stationary values and give rise to the usual Bloch equation
with renormalized $T_2(\infty)$ time and renormalized effective
magnetic field. We show that
$$
\frac{1}{T_2(\infty)} =\frac12\gamma_{\rm relax} +\gamma_{\rm cons}
+(N-1)\gamma'',
$$
where $\gamma''\geq 0$ is independent of $N$. For small ratio $r$
between the  strengths of the collective to the local couplings we
have $\gamma''=O(r^2)$. Consequently, if $r\sim N^{-1/2}$ then the
collective coupling gives a non-vanishing renormalization to the
(asymptotic) $T_2$ time, while if $r\sim N^{-1/2-\epsilon}$ (any
$\epsilon>0$) or smaller, then no collective effect is visible in
the dephasing. An interesting question is what happens for $r\sim
N^{-1/2+\epsilon}$ or larger. Then the expression for $T_2(\infty)$
suggests that the collective interaction may decrease the $T_2$ time
drastically for large $N$. However, this range of interaction
parameters is not accessible by our perturbation theory approach,
and more work in this direction is required. It is important to note
here that in order to derive our rigorous result, Theorem
\ref{thm1}, we need a strong smallness condition on all coupling
constants (see \fer{mm-2}). As explained in Section \ref{sec5}, we
expect that our result should hold for collective coupling constants
up to size $O(N^{-1/2})$ and local coupling constants of size
$O(N^0)$ (relative to the spin frequency). However, for
$r=O(N^{-1/2+\epsilon})$ we do not think that usual perturbation
theory can be applied, and a different approach should be taken.

We also examine the situation where we have two (or more) species of
spins, $A$ and $B$, each species coupled homogeneously to local and
collective reservoirs (with a single collective reservoir for both
species). We show that the $z$-component of the magnetization of
either species relaxes with single-spin relaxation time (associated
to that species). The transverse magnetization dephases following a
modified Bloch equation with time-dependent $T_2$-time and effective
magnetic field. For large times, the $T_2$-time of species $A$
approaches the limiting value
$$
\frac{1}{T_{2,A}(\infty)} = \frac12\gamma_{{\rm relax}, A} +\gamma_{{\rm cons},A}+(N_A-1)\gamma_A +N_B\gamma_B,
$$
where $N_A$ and $N_B$ are the number of spins in each class, and $\gamma_A$, $\gamma_B\geq 0$. For small ratio $r_A$, $r_B$ of the collective and local coupling constants, we have $\gamma_A=O(r_A^2)$, $\gamma_B=O(r_B^2)$. The total magnetization is the sum of that of species $A$ and $B$. It is the sum of two terms decaying (relaxing and dephasing) at different rates, and so we cannot associate to it a total relaxation time or a total dephasing time.

The effects of collective interactions between effective spins and
thermal environments, discussed in this paper,  can represent a
significant interest, for example, in NMR, MRI, and quantum
computation. The presence of energy conserving and energy exchange
collective effects can be investigated experimentally, for example,
in NMR experiments by (i) creation and controlling of collective
effects and (ii) analyzing relaxation and dephasing time-scales and
time-dependencies of magnetization as functions of characteristic
parameters.

\section{Model, single spin dynamics}

We consider $N$ non interacting spins $1/2$ coupled to local and collective bosonic heat reservoirs. The full Hamiltonian is given by
\begin{eqnarray}
H &=& -\hbar \sum_{n=1}^N\omega_n S_n^z +\sum_{n=1}^N H_{\r_n} +H_\r \label{1}\\
&& +\sum_{n=1}^N\lambda_n S_n^x \otimes\phi_\c(g_\c) +\sum_{n=1}^N\varkappa_n S_n^z\otimes\phi_\c(f_\c) \label{2}\\
&& + \sum_{n=1}^N \mu_n S_n^x\otimes\phi_n(g_n) + \sum_{n=1}^N\nu_n S_n^z\otimes\phi_n(f_n). \label{3}
\end{eqnarray}

Below we use dimensionless variables and parameters. To do so, we
introduce a characteristic frequency, $\omega_0$, typically of the
order of spin transition frequency. The total Hamiltonian, energies
of spin states, and temperature are measured in units
$\hbar\omega_0$. The frequencies of spins, $\omega_n>0$, bosonic
excitations, $\omega(k)=c|\vec{k}|$ (where $c$ is the speed of
light), the wave vectors of bosinic excitations are normalized by
$\omega_0/c$, and all constants of interactions are measured in
units $\omega_0$.  A dimemsionless time is defined as
$t\rightarrow\omega_0t$.

In \fer{2}, \fer{3}, $\omega_n>0$ is the frequency of spin $n$,
\begin{equation}
S^z = \frac{1}{2}
\left[
\begin{array}{cc}
1 & 0\\
0 & -1
\end{array}
\right]
\mbox{\qquad and\qquad}
S^x = \frac{1}{2}
\left[
\begin{array}{cc}
0 & 1\\
1 & 0
\end{array}
\right],
\label{sz}
\end{equation}
and $S^{z,x}_n$ denotes the $S^{z,x}$ of spin $n$. $H_\r$ is the
Hamiltonian of the bosonic collective reservoir,
\begin{equation}
H_\r = \int_{\rx^3} |k| a^*(k)a(k) \d^3k, \label{n2}
\end{equation}
and $H_{\r_n}$ is that same Hamiltonian pertaining to the $n$-th
individual reservoir.  For a square-integrable {\it form factor}
$h(k)$, $k\in\rx^3$,  $\phi(h)$ is given by
\begin{equation}
\phi(h) = \frac{1}{\sqrt 2}\int_{\rx^3}\left\{ h(k) a^*(k) + h(k)^* a(k)\right\} \d^3k.
\label{phi}
\end{equation}
The real numbers $\lambda_n$, $\varkappa_n$, $\mu_n$, $\nu_n$  are
coupling constants, measuring the strengths of the various
interactions as follows:
$$
\begin{array}{ll}
\lambda_n & \mbox{energy exchange collective coupling}\\
\varkappa_n &\mbox{energy conserving collective coupling} \\
\mu_n & \mbox{energy exchange local coupling}\\
\nu_n & \mbox{energy conserving local coupling}
\end{array}
$$
We introduce the maximal size of all couplings,
\begin{equation}
\alpha:=\max_n\{|\varkappa_n|, |\lambda_n|, |\mu_n|, |\nu_n|\}.
\end{equation}

The energies of the $N$ uncoupled spins are the eigenvalues  of
$H_\ss = -\sum_{n=1}^N \omega_n S_n^z$, given by
$-\frac{1}{2}\sum_{n=1}^N\omega_n\sigma_n$, where
$\sigma_n\in\{1,-1\}$. We denote by
$\varphi_\usigma=\varphi_{\sigma_1}\otimes\cdots\otimes\varphi_{\sigma_N}$
the corresponding eigenvector. {\bf Bohr energies} (energy
differences) are thus given by
\begin{equation}
e(\usigma,\utau) = -\frac{1}{2}
\sum_{n=1}^N\omega_n(\sigma_n-\tau_n). \label{4}
\end{equation}
{\bf Assumptions.}
\begin{itemize}
\item[{\bf (A)}] We consider the spin  frequencies  $\{\omega_n\}$ to be uncorrelated in the following sense:
\begin{equation}
\mbox{
\it If $e(\usigma,\utau)=e(\usigma',\utau')$ then $\sigma_n-\tau_n = \sigma'_n-\tau'_n$ for all $n$.}
\label{assumption}
\end{equation}
In particular, we do not allow any of the $\omega_n$ to be the equal. However, we can describe a {\it homogeneous magnetic field} within the constraint \fer{assumption} by considering a distribution $\omega_n=\omega +\delta\omega_n$ for some fluctuation $\delta\omega_n$ having, say, uniform distribution in some interval. Then assumption (A) is satisfied almost surely. In view of such fluctuations, relation \fer{assumption} is reasonable from a physical point of view, and its mathematical advantage is that it breaks permutation symmetry and hence reduces the degeneracies of the energies $e$.

\item[{\bf (B)}] The smallest gap between different Bohr energies of the  non-interacting spin energies \fer{4} is
\begin{equation}
\Delta =\frac{1}{2}\min_{m_n,m'_n}\left\{ \left| \sum_{j=1}^N
\omega_n(m_n-m_n')\right|\right\}\backslash\{0\}, \label{mm-1}
\end{equation}
where the minimum is taken over sequences $m_n, m'_n\in\{-2,0,2\}$.
As our approach is based on perturbation theory of Bohr energy
differences,  their displacement under interaction, which is of size
$N^2\alpha^2$, should be small relative to $\Delta$,
\begin{equation}
N^2\lambda^2 <\!\!< \Delta.
\label{mm-2}
\end{equation}
For $\omega_n=\omega$ constant, we have $\Delta=\hbar\omega$.  Hence
for a homogeneous magnetic field $\omega_n=\omega+\delta\omega_n$
with small fluctuation $\langle\delta\omega_n\rangle/\omega<\!\!<1$,
we have $\Delta =\omega +O(\langle\delta\omega\rangle)$, and the
r.h.s. of \fer{mm-2} is independent of $N$. Condition \fer{mm-2} is
a serious restriction on the coupling strength for large systems
(big $N$). Our analysis uses this condition in several technical
estimates of remainder terms, stemming from perturbation theory (see
also \cite{MSB1}). However, it is seen from physical considerations,
presented in Section \ref{sec5}, that the true condition should read
$\alpha^2_cN<\!\!<\omega$ and $\alpha_\ell <\!\!< \omega$, where
$\alpha_c$ and $\alpha_\ell$ are the sizes of collective and local
coupling constants, and $\omega$ is the typical frequency of a spin.

\item[{\bf (C)}] Regularity of form factors: denote by $h$ any of the functions $f_\c, g_\c, f_n,g_n$ in the Hamiltonian $H$. Let $r\geq 0$, $\Sigma\in S^2$ be the spherical coordinates of $\rx^3$. Then $h(r,\Sigma) =r^p\e^{-r^m}h'(\Sigma)$, with $p=-1/2+n$, $n=0,1,2,\ldots$ and $m=1,2$, and where $h'$ is any angular function. (Less restrictive requirements on $h$ are necessary only \cite{MSB1}, but they are more technical to describe, so we restrict our attention to $h$ satisfying this condition. This family of form factors contains the usual physical ones \cite{PSE}.)

\end{itemize}

\noindent
Given $e$, \fer{4}, the number
\begin{equation}
N_0(e) = \{ n\ :\ \sigma_n=\tau_n\ \mbox{ for any $(\usigma,\utau)$ with $e(\usigma,\utau)=e$}\}
\label{N0}
\end{equation}
depends on $e$ alone, and the number of different configurations $(\usigma,\utau)$ with constant value \fer{4} is $2^{N_0(e)}$. There are $2^{N_0(e)}$ elements $\scalprod{\varphi_\usigma}{\rho_\ss \varphi_\utau}$ of the (reduced) density matrix of the spins with fixed value \fer{4}. As shown in \cite{MSB1}, these elements evolve in time jointly, and independently of elements associated with any other value of \fer{4}.

We consider unentangled initial states
$$
\rho_0= \rho_{\s_1}\otimes\cdots\otimes \rho_{\s_N}\otimes\rho_{\r_1}\otimes\cdots\otimes\rho_{\r_N}\otimes\rho_\r,
$$
where $\rho_{\s_j}$ are arbitrary single spin states, and $\rho_{\r_j}$, $\rho_\r$ are thermal equilibrium states of single reservoirs, all at temperature $T=1/\beta>0$.

The reduced density matrix $\rho^{(j)}_t$ of spin $j$ is given by
$$
\rho_t^{(j)} = {\rm Tr}^{(j)}\  \e^{-\i t H} \rho_0 \e^{\i tH},
$$
the trace being taken over all spins $n\neq j$ and over all reservoirs.

Let $A_j$ be an observable of the $j$-th spin, and denote its dynamics by
$$
\av{A_j}_t = {\rm Tr}\rho^{(j)}_t A_j,
$$
where the trace is taken over the space of $S_j$. Our goal is to find a representation of $\av{A_j}_t$. For a square integrable form factor $h(k)=h(|k|,\Sigma)$ (spherical coordinates of $\rx^3)$, the {\bf spectral density} of the reservoir associated to $h$ is given by
\begin{equation}
J_h(\omega) = \pi\omega^2 \int_{S^2}|h(\omega,\Sigma)|^2 \d\Sigma.\  \footnote{Let  $C_h(t)=\frac12 [\langle\phi(h)\e^{\i tH_\r}\phi(h)\e^{-\i tH_\r}\rangle_\beta+ \langle\e^{\i tH_\r}\phi(h)\e^{-\i tH_\r}\phi(h)\rangle_\beta]$ be the symmetrized correlation function of a reservoir in thermal equilibrium at temperature $T=1/\beta$, with $H_\r$ and $\phi(h)$ given in \fer{n2} and \fer{phi}. The Fourier transform $\widehat C_h(\omega)=\int_0^\infty \e^{-\i \omega t}C(t)\d t$, $\omega\geq 0$, is related to the spectral density by
$$
{\rm Re}\, \widehat C_h(\omega) = J_h(\omega)\coth(\beta \omega/2).
$$
}
\label{specdens}
\end{equation}
Decay rates are given by coupling constants squared times $J_h(\omega)\coth(\beta\omega/2)$
 at values $\omega$ corresponding to Bohr frequencies of the spin system. Energy-conserving processes are associated with the Bohr frequency $\omega=0$, and since $\coth(\beta\omega/2)\sim \omega^{-1}$ as $\omega\sim 0$, we introduce
\begin{equation}
\widetilde J_h(0) = \lim_{\omega\rightarrow 0_+} \frac{J_h(\omega)}{\omega}.
\label{tildeJ}
\end{equation}
Define the quantities
\begin{eqnarray}
b_j&=& \frac{1}{4}\frac{\e^{\beta\hhbar\omega_j}}{\e^{\beta\hhbar\omega_j}-1}  \left\{ \lambda^2_j J_{g_\c}(\hhbar\omega_j)+\mu^2_j J_{g_j}(\hhbar\omega_j)\right\}\label{bj}\\
c_j&=&\e^{-\beta\hhbar\omega_j}\label{cj}\\
Z_{\beta,j}&=&\e^{-\beta\hhbar\omega_j/2}+\e^{\beta\hhbar\omega_j/2}\\
X_j &=&\frac{1}{8\pi}{\rm P.V.}\int_{\rx} \frac{\lambda^2_j J_{g_\c}(|u|)+\mu^2_j J_{g_j}(|u|)}{u+\hhbar\omega_j}\coth(\beta|u|/2)\d u\label{Xj}\\
Y_j&=& \frac{1}{8}\left\{ \lambda^2_j J_{g_\c}(\hhbar\omega_j) + \mu^2_j J_{g_j}(\hhbar\omega_j)\right\} \coth(\beta\hhbar\omega_j/2) +\frac{1}{2\beta}\big\{ \varkappa^2_j\widetilde J_{f_\c}(0) +\nu^2_j\widetilde J_{f_j}(0)\big\}\ \ \ \ \ \ \ \ \ \label{Yj}
\end{eqnarray}
With this notation in place we have the following result.
\begin{thm}[Dynamics of single spin]
\label{thm1}
For any observable $A_j$ of spin $j$, $j=1,\ldots,N$, and $t\geq 0$, we have
\begin{eqnarray}
\lefteqn{
\av{A_j}_t = Z_{\beta,j}^{-1}\ {\rm Tr\ }\e^{-\beta H_{\s_j}}A_j}\label{f0}\\
&&  +\e^{-tb_j(c_j+1)}\left\{ [\rho_0^{(j)}]_{11} -\frac{1}{\e^{-\beta\hhbar\omega_j}+1}\right\}
\big([A_j]_{11} -[A_j]_{22}\big)\qquad \label{f00}\\
&& +\e^{\i t( -\hhbar\omega_j + X_j +\i Y_j)}{\cal C}_j(N,t)\  [\rho_0^{(j)}]_{21} \ [A_j]_{12}\label{f1}\\
&& +\overline{\e^{\i t(-\hhbar\omega_j + X_j +\i Y_j)} {\cal C}_j(N,t)\ [\rho_0^{(j)}]_{21}}\  [A_j]_{21}\label{f2}\\
&&+ O(\alpha^2).
\label{5}
\end{eqnarray}
\end{thm}

\bigskip
\noindent
The quantity ${\cal C}_j(N,t)$ involves the interaction parameters and the initial condition of all spins other than $j$,
\begin{eqnarray}
{\cal C}_j(N,t) &=& \prod_{l\neq j} \left\{ \left[ \e^{\i tz_l^+} -\e^{\i tz_l^-}\right] \frac{1+c_l\alpha_l}{1+c_l\alpha^2_l} \big(\alpha_l+ [\rho_0^{(l)}]_{11}(1-\alpha_l)\big) +\e^{\i tz_l^-}\right\} \label{calcj}\\
z^\pm_l&=& \frac{1}{2}\left\{ \i b_l(1+c_l) \pm \sqrt{-b^2_l(1+c_l)^2 +4a_l[a_l-\i b_l(1-c_l)]}\right\} \label{zj}\\
\alpha_l  &=& 1+\i\frac{z_l^+-a_l}{b_lc_l}
\end{eqnarray}
In \fer{zj}, the square root is the principal value (cut on the negative real axis), $b_l, c_l$ are given in \fer{bj}, \fer{cj}, and
\begin{equation}
a_l= -\frac{1}{2} \varkappa_l^2 \, {\rm P.V.} \int_{\rx^3}\frac{|f_{\c}(p)|^2}{|p|}\d^3p.\label{aj}
\end{equation}

\medskip
\noindent
{\bf Discussion of the factor ${\cal C}_j(N,t)$.\ }

\smallskip
\noindent
Clearly ${\cal C}_j(N,0)=1$. For vanishing energy-conserving collective coupling, $\varkappa_l=0$ (all $l$), we have ${\cal C}_j(N,t)=1$ for all $t\geq 0$. This follows from $a_l=0$, $z_l^+ = \i b_l(1+c_l)$, $z_l^-=0$ and $\alpha_l=-1/c_l$.

As soon as the collective energy-conserving coupling is switched on, $\varkappa_l\neq 0$, the analysis of ${\cal C}_j(N,t)$ is difficult. The factors in the product \fer{calcj} decay with rate at least
\begin{equation}
\gamma_j=\min\{\Im z_j^+,\Im z_j^-\},
\label{gammajj}
\end{equation}
so for a homogeneous system (each factor the same) we have the estimate $|{\cal C}(N,t)|\leq C_N \, \e^{-\gamma (N-1)t}$, with
\begin{equation}
\gamma=\min_j\gamma_j.
\label{gamma}
\end{equation}
Of course, ${\cal C}_j$ does not depend on $j$ anymore.  This estimate says that ${\cal C}$ decays in time with rate $\gamma(N-1)$, but we have a prefactor depending on $N$. We have the upper bound $C_N\leq \e^{(N-1)c'}$, for
\begin{equation}
c'= \ln \left\{2\left|\frac{1+c\alpha}{1+c\alpha^2}(\alpha +[\rho_0]_{11}(1-\alpha)\right|+1\right\} >0.
\label{c'}
\end{equation}
We know that for $t=0$ the true upper bound on $|{\cal C}(N,t)|$ corresponds to $C_N=1$, but this does not mean at all that $|{\cal C}(N,t)|\leq \e^{-\gamma (N-1)t}$.
The estimate
\begin{equation}
|{\cal C}(N,t)|\leq \e^{(N-1)[-\gamma t+c']},\qquad \mbox{with\ \ $\gamma=\min_j\gamma_j$, see \fer{gammajj}}
\label{bndcalcj}
\end{equation}
shows that $|{\cal C}|$ decays to $1/2$ (half of its initial value) no later than at time $\gamma^{-1}[\frac{\ln 2}{N-1}+c']$. We thus call
\begin{equation}
\gamma'=\gamma\left[\frac{\ln 2}{N-1} +c'\right]^{-1} \qquad (\gamma'\approx\gamma/c'\mbox{\ \ for large $N$})
\label{preratio}
\end{equation}
the decay rate of of $|{\cal C}(N,t)|$. Let us examine this decay rate in the two cases where
\begin{equation}
r=\frac{|a_l|}{b_l}\sim\frac{\varkappa^2_l}{\lambda_l^2+\mu_l^2}
\label{ratio}
\end{equation}
is either very small or very close to one. \fer{ratio} very small corresponds to the situation where the collective interactions ($\varkappa\approx\lambda$) are much smaller than the local ones ($\mu\approx\nu$). The situation where \fer{ratio} is unity describes very large collective coupling relative to the local ones.

\medskip
{\it For small collective coupling}, $r\approx 0$, we obtain $\gamma'\sim{\rm const.} r \varkappa$, where ${\rm const.}$ does not depend on $N$. In the limit $r\rightarrow 0$ we get $\gamma'=0$, which is the correct behaviour as we have seen above (${\cal C}=1$ in this setting, no decay).

{\it For large collective coupling}, $r\approx 1$, we obtain  $\gamma'\sim{\rm const.}b$, where ${\rm const.}$ does not depend on $N$ and $b$ is given in \fer{bj}.


\bigskip
\noindent
{\bf Comparison to exactly solvable model.\ }

\smallskip
\noindent
If the spins interact with the reservoirs only through energy conserving channels, then $\lambda_n=\mu_n=0$ in \fer{2}, \fer{3}. This model is {\it exactly solvable}. By proceeding as in \cite{MSB1} (``Resonance theory of decoherence and thermalization'', proof of Proposition 7.4), one finds the following {\it exact formula} for the evolution of the reduced density matrix element of a single spin. For simplicity of notation, we take all $\varkappa_n$ to be constant $\varkappa_c$ (collective) and all $\nu_n$ constant $\nu_\ell$ (local). We also take all local form factors equal ($f_\ell$) and all collective ones too ($f_c$). Then we find
\begin{equation}
[\rho_t^{(j)}]_{21}=  [\rho_0^{(j)}]_{21} \ \e^{-\i\omega_j t}\ \e^{-\nu_\ell^2\Gamma_\ell(t) -\varkappa_c^2\Gamma_c(t)} \sum_{\sigma_k, \, k\neq j} \ \prod_{l, \, l\neq j}^N [\rho_0^{(l)}]_{\sigma_l\sigma_l} \e^{-2\i\sigma_l\varkappa^2_cS_c(t)}.
\label{exact1}
\end{equation}
The sum is over $\sigma_k=\pm 1/2$, $k=1,\ldots,N$, $k\neq j$, where $\sigma_j=1/2$ corresponds to the energy eigenstate $\varphi_1=[1 \ 0]^T$ of $S^z$ (see \fer{sz}). The decoherence functions and Lamb shift are given by
\begin{eqnarray}
\Gamma(t) &=&\int_{\rx^3} |f(k)|^2 \coth(\beta|k|/2) \frac{\sin^2(|k|t/2)}{|k|^2} \d^3k\\
S(t) &=& -\frac12 \int_{\rx^3}|f(k)|^2 \frac{|k|t -\sin(|k|t)}{|k|^2} \d^3k.
\end{eqnarray}
Of course, the populations are time-independent in this model, $[\rho^{(j)}_t]_{ll} =[\rho^{(j)}_0]_{ll}$ for all $t\geq0$, $l=1,2$. The time-dependence in the exponentials in \fer{exact1} becomes linear for large times, $\Gamma(t)\rightarrow t\widetilde J(0)$ and $\varkappa^2_cS(t)\rightarrow t a$ as $t\rightarrow\infty$, where $\widetilde J(0)$ and $a$ are given in \fer{tildeJ} and \fer{aj} with $\varkappa_l$ replaced by $\varkappa_c$ (see \cite{MSB1}).

It is not hard to see that upon the replacements $\Gamma(t)\mapsto t\widetilde J(0)$ and $\varkappa^2_cS(t)\mapsto ta$ the exact formula \fer{exact1} coincides precisely with expression \fer{f0}-\fer{f2} for $A$ s.t. $[A]_{11}=[A]_{22}=[A]_{21}=0$, $[A]_{12}=1$ (so that $\langle A\rangle_t=[\rho_t]_{21}$). The factor ${\cal C}_j(N,t)={\cal C}(N,t)$ is thus identified with the sum of the product in \fer{exact1}. If all spins are initially in the same state, characterized by the population probability $0\leq p\leq 1$ for the state with $\sigma=1/2$, we obtain
$$
{\cal C}(N,t) = \left[p\e^{-\i a t} +(1-p)\e^{\i a t}\right]^{N-1},\qquad\mbox{ with $a$ given in \fer{aj}.}
$$
Clearly $|{\cal C}(N,t)|\leq 1$ for all times and all $N$. Also, for all $n\in\mathbb Z$, we have $|{\cal C}(N,n\pi a^{-1})|=1$ and $|{\cal C}(N,(n+\frac{1}{2})\pi a^{-1})|=|1-2p|^{N-1}\approx 0$ for $N$ large and $p\neq 0,1$. Therefore the factor ${\cal C}(N,t)$ oscillates in size between zero and one, with frequency $|a|/\pi$ proportional to the square of the energy-conserving collective coupling $\varkappa_c^2$.

\section{Evolution of single spins and of magnetization}

\subsection{Single spin relaxation and dephasing times}
\label{ss1}

 The term on the r.h.s. of \fer{f0} is the equilibrium average at temperature $T=1/\beta$. From \fer{f00} we obtain the relaxation rate of spin $j$, namely $\gamma_{{\rm relax},j}=b_j(c_j+1)$. The single spin relaxation rate is
\begin{equation}
\gamma_{{\rm relax},j}= 1/\tau_{{\rm relax},j} = \frac14 \coth(\beta \hhbar\omega_j/2) \left\{ \lambda^2_j J_{g_\c}(\hhbar\omega_j)+\mu^2_j J_{g_j}(\hhbar\omega_j)\right\}.
\label{gammaj}
\end{equation}
The single-spin relaxation time depends on the local ($\mu_j$) and collective ($\lambda_j$) couplings in the same manner: {\it In the relaxation process, the collective reservoir acts as a local reservoir.}


Next we consider the dephasing time determined by \fer{f1}, \fer{f2}. There are two contributions to the time decay. One comes from spin $j$ itself and is given by $Y_j$, the other one comes from all other spins than $j$ and is given by ${\cal C}_j$. One sees from \fer{zj} that $\Im z_l^\pm\geq 0$, and that $\min\{\Im z_l^+,\Im z_l^-\}=0 \Leftrightarrow a_lb_l=0$. It follows that if the energy conserving collective coupling and at least one of the energy-exchange couplings (local or collective) do not vanish (so that $a_lb_l\neq 0$), then we have \fer{bndcalcj} with $\gamma>0$.
The single-spin dephasing rate is thus $Y_j+\gamma'$ which we can write as
\begin{equation}
\gamma_{{\rm deph},j} = \frac{1}{2}\gamma_{{\rm relax},j}+\gamma_{{\rm cons},j}+\gamma',
\label{mm-4}
\end{equation}
where
\begin{equation}
\gamma_{{\rm cons},j} = \frac{1}{2\beta}\big\{ \varkappa_j^2 \widetilde J_{f_c}(0)+\nu_j^2\widetilde J_{f_j}(0)\big\}
\label{mm-5}
\end{equation}
is a contribution coming purely from the energy-conserving interactions, in which the local and collective couplings play the same role. The last term in expression \fer{mm-4} is due to the presence of the $N-1$ spins other than the considered one. As we have seen after \fer{ratio}, if the collective coupling is small ($r\approx 0$), then $\gamma'\sim \varkappa^2r<\!<\!\varkappa^2$ and hence the last term in \fer{mm-4} is negligible. If the collective coupling is large ($r\approx 1$), then $\gamma'\sim b\sim \gamma_{\rm relax,j}$.

\medskip
{\bf Conclusions.\ } $\bullet$ The single-spin relaxation rate is the sum of two contributions from the local and the collective energy-exchange interactions \fer{gammaj}. The collective term has the same form as the local term, and the presence of all other spins does not influence the single spin relaxation rate.

$\bullet$  The single-spin dephasing rate has three contributions \fer{mm-4}. One is half the relaxation rate (exchange interactions), one comes from energy conserving interactions (local and collective), and a third term which is due to the presence of all other spins. That last term ($\gamma'$) is negligible for small collective coupling, and renormalizes the dephasing rate for strong collective couplings by an amount independent of the number of spins.


\subsection{Evolution of magnetization}
Let
$$
\vec S =
\left[
\begin{array}{c}
S^x\\
S^y\\
S^z
\end{array}
\right]
$$
be the total magnetization vector, where $S^{x,y,z}=\sum_{j=1}^N S^{x,y,z}_j$. It is convenient to introduce the complex (non-hermitian) observable
$$
S_j^- = S_j^x -\i S^y_j.
$$
We use Theorem \ref{thm1} with $A_j=S_j^{x,y,z}$ to obtain
\begin{eqnarray}
\av{S_j^z}_t &=&\frac{1}{2}\tanh(\beta\hhbar\omega_j/2)
 [1-\e^{-t/\tau_{{\rm relax},j}}] +\e^{-t/\tau_{{\rm relax},j}}\av{S_j^z}_0
 +O(\alpha^2),
\label{3.8}\\
\langle S^-_j\rangle_t &=&\e^{\i t (-\hhbar\omega_j +X_j+ \i Y_j)}
{\cal C}_j(N,t) \langle S_j^-\rangle_0 +O(\alpha^2). \label{3.9}
\end{eqnarray}

\medskip
\noindent {\bf Purely local coupling.\ } In the absence of
collective  coupling ($\lambda_n=0=\varkappa_n$), the above
equations simplify to
\begin{eqnarray}
\av{S_j^z}_t &=&\frac{1}{2}\tanh(\beta\hhbar\omega_j/2)
[1-\e^{-t/\tau_{{\rm relax},j}}] +\e^{-t/\tau_{{\rm
relax},j}}\av{S_j^z}_0 +O(\alpha^2),
\label{3.8'}\\
\langle S^-_j\rangle_t &=&\e^{\i t (-\hhbar\omega_j +X_j+\i Y_j)}
\langle S_j^-\rangle_0 +O(\alpha^2). \label{3.9'}
\end{eqnarray}
where $\tau_{{\rm relax},j}$ is given by $1/\gamma_{{\rm
relax},j}$,. \fer{gammaj} with $\lambda_j=0$, $X_j, Y_j$ are given
in \fer{Xj}, \fer{Yj} with $\lambda_j=\mu_j=0$. The factor ${\cal
C}_j(N,t)$ equals $1$ (as discussed after \fer{aj}).

\subsubsection{Homogeneous magnetic field}

In this section we derive the evolution of the magentization vector in a homogeneous magnetic field, characterized by $\omega_j=\omega+\delta\omega_j$ with $\delta\omega_j\rightarrow 0$ (see also assumption (B) after \fer{mm-2}). This is the description of an elementary volume of many spins sitting in a magnetic field with gradient much smaller than the size of the elementary volume.

\medskip
We consider all spins initially in the same state. We take all local couplings to be the same, i.e., $g_j=g_\ell$ and all collective couplings to be the same, $f_j=f_{\rm c}$, and all coupling constants independent of $j$. In this limit, we have in formulas \fer{3.8}, \fer{3.9}
$$
\omega_j=\omega,  \quad X_j=X, \quad Y_j=Y,
$$
where $\tau_{{\rm relax}}$, $X$, $Y$ are given in \fer{gammaj}, \fer{Xj}, \fer{Yj}, with $\omega_j, f_j, g_j$ and all coupling constants replaced by their constant values, in particular,
\begin{equation}
\gamma_{{\rm relax}} =1/\tau_{{\rm relax}} = \frac14  \coth(\beta \hhbar\omega/2) \left\{ \lambda^2 J_{g_\c}(\hhbar\omega)+\mu^2 J_{g_\ell}(\hhbar\omega)\right\}.
\label{102}
\end{equation}
Furthermore, we have ${\cal C}_j(N,t)={\cal C}(N,t)$, with (see \fer{calcj})
\begin{eqnarray}
{\cal C}(N,t)&=& [{\cal D}(t)]^{N-1}\label{124.1}\\
{\cal D}(t) &=& \left[\e^{\i tz^+} -\e^{\i tz^-}\right] \frac{1+c\alpha}{1+c\alpha^2} \big(\alpha+ [\rho_0]_{11}(1-\alpha)\big) +\e^{\i tz^-}.
\label{124}
\end{eqnarray}
We sum equations \fer{3.8} and \fer{3.9} over $j$ to obtain (dropping the $O(\alpha^2)$ terms)
\begin{eqnarray}
\langle S^z\rangle_t &=& \frac{N}{2}\tanh(\beta\hhbar\omega/2) [1-\e^{-t/\tau_{{\rm relax}}}] +\e^{-t/\tau_{{\rm relax}}}\av{S^z}_0 \label{100}\\
\langle S^-\rangle_t &=&\e^{\i t (-\hhbar\omega +X+\i Y)} [{\cal D}(t)]^{N-1} \langle S^-\rangle_0.\label{101}
\end{eqnarray}
It is clear that \fer{100} is the integrated version of the Bloch equation
\begin{equation}
\frac{\d}{\d t}\av{S^z}_t = -\frac{1}{\tau_{{\rm relax}}}\big[\av{S^z_j}_t -\frac N2 \tanh(\beta\hhbar\omega/2)\big]
\label{++}
\end{equation}
corresponding to the homogeneous magnetic field $\vec B=B_z\vec e_z = -\hhbar\omega \vec e_z$, with relaxation time
$$
T_1=\tau_{{\rm relax}}=1/\gamma_{\rm relax},
$$
see \fer{102}. The Bloch equation for the {\it transverse magnetization} would read
\begin{equation}
\frac{\d}{\d t} \langle S^-\rangle_t = -\frac{1}{T_2} \langle S^-\rangle_t +\i B_z\langle S^-\rangle_t.
\label{104}
\end{equation}
However the true evolution, \fer{101}, is not of this form. By differentiating \fer{101} we obtain
\begin{equation}
\frac{\d}{\d t} \langle S^-\rangle_t = -\Gamma(t) \langle S^-\rangle_t +\i B(t)\langle S^-\rangle_t,
\label{105}
\end{equation}
with
\begin{eqnarray}
\Gamma(t) &=& \frac12\gamma_{\rm relax} +\gamma_{\rm cons} -(N-1) \,
{\rm Re}\frac{\d}{\d t}\ln {\cal D}(t) \label{110},\\
B(t) &=& -\hhbar \omega +X+(N-1) \, {\rm Im} \frac{\d}{\d t} \ln {\cal D}(t),\label{111}
\end{eqnarray}
where $\gamma_{\rm relax}$ is the single-spin relaxation rate \fer{gammaj}  and $\gamma_{\rm cons}$ is the single-spin dephasing rate due to the energy-conserving interactions \fer{mm-5}.

Comparing \fer{105} with \fer{104} leads us to the  identification
of a {\it time-dependent dephasing time $T_2=1/\Gamma(t)$} and {\it
a time-dependent effective magnetic field $B_z=B(t)$}.

The deviation of the true equation of evolution from the Bloch
equation is given by the terms $\d/\d t \ln {\cal D}(t)$ in
\fer{110}, \fer{111}. We now estimate the size of this term for weak
collective coupling, where $r$ is small, see \fer{ratio}. It is not
hard to see that
\begin{equation}
\left| \frac{\dot{\cal D}(t)}{{\cal D}(t)}\right|\leq C|r|, \qquad
\lim_{t\rightarrow\infty} \frac{\dot{\cal D}(t)}{{\cal D}(t)} = \i
z^-=4\i br\tanh(\beta \omega/2) +O(r^2), \label{**}
\end{equation}
for a constant $C$ independent of $t$ (and $N$).

\medskip
{\bf Conclusions.\ } The Bloch equation for the total magnetization
(homogeneous magnetic field) holds with relaxation time $T_1$ given
by the single-spin relaxation rate \fer{gammaj} (no influence of the
other spins). The total magnetization dephases with a time-dependent
$T_2$-time, $T_2=T_2(t)$. We have $1/T_2(t)=\frac12\gamma_{\rm
relax}+\gamma_{\rm cons} +(N-1){\rm Re} Y'(t)$, see also
\fer{gammaj}, \fer{mm-5}. The time-dependent part $Y'(t)$ stems from
the collective interaction. For weak collective interaction, $r$
small (see \fer{ratio}), we have $|Y'(t)|\leq C|r|$ (all times).

The term $(N-1){\rm Re}\frac{\d}{\d t}{\cal D}(t)$ in \fer{110} is
$O(Nr^2)$  for large times, see \fer{**}. If $r$ is of the order
$1/\sqrt N$ then this is of order one, and the collective
interaction gives time-dependent modification of the dephasing time
$T_2$ with an asymptotically renormalized value
$$
1/T_2(\infty) = \frac12\gamma_{\rm relax}+\gamma_{\rm cons}+(N-1){\rm Im }
 z^-,\qquad \mbox{${\rm Im}z^-=O(r^2)$}.
$$
If $r$ is smaller than $N^{-1-\epsilon}$ (any $\epsilon>0$)  then
the collective interaction has no effect in \fer{110}, \fer{111} and
the Bloch equation for transversal relaxation holds with $T_2$ the
single-spin dephasing time $[\gamma_{\rm relax}/2+ \gamma_{\rm
cons}]^{-1}$. For larger collective interaction we may get large
corrections to the Bloch equation, since the last terms in
\fer{110}, \fer{111} may become large (big $N$). This regime does
not enter the present perturbative setup, and more work on this
issue is needed.

Note that in any event, since ${\rm Im}z^-\geq 0$, the collective
interactions can only  accelerate the dephasing process.

\subsubsection{Multi-species inhomogeneity}

Consider the situation where $N$ spins are grouped into two (or
more) classes $A$ and $B$.  We describe the situation where within
each class, the spins are homogeneous. We have two magnetic fields
$\omega_A$, $\omega_B$, two sets of coupling constants ($\lambda_A$,
$\lambda_B$ etc), two sets of form factors ($g_{{\rm c},A}$,
$g_{{\rm c},B}$, $g_{A}$, $g_{B}$ etc). Let $N_A$ and $N_B$ be the
relative sizes,
$$
N_A+N_B=N.
$$
If spin $j$ belongs to class $A$, then \fer{calcj} becomes
\begin{equation}
{\cal C}_j(N,t) = \left[{\cal D}_A(t)\right]^{N_A-1}\left[{\cal D}_B(t)\right]^{N_B},
\label{120}
\end{equation}
with ${\cal D}_A(t), {\cal D}_B(t)$ given as in \fer{124} for species $A,B$. Let
$$
\vec S_A = \sum_{j\  {\rm{in\,  class}} \, A} \vec S_j
$$
and, correspondingly, for the three components of this vector. We
sum \fer{3.8}  and \fer{3.9} over all indices of spins belonging to
class $A$ to obtain
\begin{eqnarray}
\langle S^z_A\rangle_t &=& \frac{N_A}{2}\tanh(\beta \omega_A/2) [1-\e^{-t/\tau_{{\rm relax},A}}] +\e^{-t/\tau_{{\rm relax},A}}\langle S^z_A\rangle_0 +O(\alpha^2)\label{3.8''}\\
\langle S^-_A\rangle_t &=& \e^{\i t(-\hhbar \omega_A +X_A +\i Y_A)} [{\cal D}_A(t)]^{N_A-1} [{\cal D}_B(t)]^{N_B} \langle S^-_A\rangle_0 +O(\alpha^2).
\label{3.9''}
\end{eqnarray}
Hence class $A$ relaxes with single-spin relaxation time $\tau_{{\rm relax},A}$ according to the usual Bloch equation \fer{**}. For the transverse magnetization we obtain again a modified Bloch equation with time-dependent relaxation time and effective magnetic field,
\begin{equation}
\frac{\d}{\d t}\langle S^-_A\rangle_t = -\Gamma_A(t) \langle S_A^-\rangle_t +\i B_A(t)\langle S^-_A\rangle_t +O(\alpha^2),
\label{121}
\end{equation}
with
\begin{eqnarray}
\Gamma_A(t) &=& \frac12\gamma_{{\rm relax},A}+\gamma_{{\rm cons},A} -(N_A-1) \,{\rm Re}\frac{\d}{\d t}\ln {\cal D}_A(t) -N_B\,{\rm Re}\frac{\d}{\d t}\ln {\cal D}_B(t) \qquad\label{110a}\\
B_A(t) &=& -\hhbar \omega_A +X_A+(N_A-1) \, {\rm Im} \frac{\d}{\d t} \ln {\cal D}_A(t) +N_B {\rm Im}\frac{\d}{\d t}{\cal D}_B(t).\label{111a}
\end{eqnarray}
As in the previous paragraph, we see that for weak collective coupling and large times, $\Gamma_A(t)$ converges to $\frac12\gamma_{{\rm relax},A}+\gamma_{{\rm cons},A}- (N_A-1){\rm Im}z^-_A$, and $B_A(t)$ converges to $-\hhbar\omega_A+X_A+(N_A-1){\rm Im} z^-_A$. We thus obtain the (asymptotic) dephasing rate for species $A$,
\begin{equation}
\gamma_{{\rm deph},A}(\infty) =  \frac12\gamma_{{\rm relax},A}+\gamma_{{\rm cons},A}+ (N_A-1){\rm Im }z^-_A +N_B{\rm Im}z^-_B.\ \footnote{
A straightforward generalization to $s$ species $A_1,\ldots, A_s$ with sizes $N_{A_1}+\cdots+N_{A_s}=N$ gives the transverse relaxation (i.e., dephasing) rates
$$
\gamma_{{\rm deph},A_j}(\infty)=  \frac12\gamma_{{\rm relax},A_j}+\gamma_{{\rm cons},A_j}+(N_{A_j}-1){\rm Im}z_{A_j} + \sum_{k\neq j} N_{A_k} {\rm Im} z_{A_k},
$$
{}for $j=1,\ldots,s$. We conclude that the relaxation rate of each species is a single-spin relaxation rate, while the dephasing contains collective effects. In particular, the dephasing rate of class $A_j$ depends on all other classes.}
\label{125}
\end{equation}
Recall again that for small collective interaction, ${\rm Im}z_{A,B}^-=O(r^2_{A,B})$, \fer{ratio}.

\medskip
{\bf Conclusions.\ } The $z$-component of the total magnetization of each species $A$ and $B$ evolves according to the Bloch equation \fer{++} with single-spin relaxation rates $\gamma_{{\rm relax},A}$ and $\gamma_{{\rm relax},B}$, \fer{gammaj}.

The transverse total magnetization of species $A$ evolves according to a modified Bloch equation \fer{121} (similarly for $B$). The dephasing time becomes time-dependent \fer{110a}, and takes the value $T_{2,A}(\infty)=1/\gamma_{{\rm deph},A}(\infty)$, \fer{125} for large times and small collective coupling.

The total magnetization is the sum of that of species $A$ and $B$, $\langle S\rangle_t = \langle S_A\rangle_t +\langle S_B\rangle_t$. The $z$-component relaxes as a sum of two exponentially decaying quantities with different rates (corresponding to $A$ and $B$). Therefore we cannot associate to it a total a single decay rate.

The total transverse magnetization is the sum of that of species $A$ and $B$. Each contribution evolves according to the modified Bloch equation. For large times, the dephasing time approaches a renormalized constant value. Being again a sum of two terms decaying at different rates, the total transverse magnetization does not have a single decay rate.

\section{Proof of Theorem \ref{thm1}}

We set $j=1$ in this proof (the case of general $j$ is obtained merely by a change in notation). Following the method developed in \cite{MSB1}, the dynamics of $A_1$ is represented as
\begin{equation}
\av{A_1}_t = \scalprod{\psi_0}{B_1\cdots B_N \e^{\i tK} A_1\Omega_{\ss}\otimes\Omega_{\rr}}.
\label{6}
\end{equation}
 The scalar product on the r.h.s. is that of the GNS Hilbert space (``doubled space''). Here, $\Omega_\ss=\Omega_{\s_1}\otimes\cdots \otimes\Omega_{\s_N}$, $\Omega_{\rr}=\Omega_{\r_1}\otimes\cdots\otimes\Omega_{\r_N}\otimes\Omega_\r$ and $\Omega_{\s_j}$ is the trace state of $S_j$, $\scalprod{\Omega_{\s_j}}{A\Omega_{\s_j}} = \frac12 ([A]_{11}+[A]_{22})$, and $\Omega_{\r_j}$ are reservoir equilibrium states at temperature $T=1/\beta$.

The $B_j$ are unique operators (in the commutant of the algebra of observables of spin $j$) satisfying
\begin{equation}
\psi_{\s_j}=B_j\Omega_{\s_j},
\label{Bj}
\end{equation}
where $\psi_{\s_j}$ is the initial state of $S_j$.

The operator $K$ is the Liouville operator acting on all spins and all reservoirs, satisfying
\begin{equation}
K\Omega_\ss\otimes\Omega_\rr=0.
\label{kernel}
\end{equation}
Its explicit form is easily written down (even though it is somewhat lengthy, see \cite{MSB1}) The main property is the representation
\begin{equation}
P_\rr\e^{\i tK}P_\rr =\sum_{e,s}\e^{\i t\varepsilon_e^{(s)}} Q_e^{(s)} +O(\alpha^2 \e^{-\gamma t}),
\label{7}
\end{equation}
where $P_\rr=|\Omega_\rr\rangle\langle\Omega_\rr|$ projects out all degrees of freedom of the reservoirs. The sum runs over all $e$ of the form \fer{4}, i.e., eigenvalues of the operator
\begin{equation}
L_\ss =H_\ss\otimes\bbbone_\ss - \bbbone_\ss\otimes H_\ss
\end{equation}
acting on $\cx^{2N}\otimes{\cx}^{2N}$ (which are also the
eigenvalues  of $K$ with $\alpha=0$). For each $e$ fixed, $s$
indexes its splitting into $\varepsilon_e^{(s)}$, $1\leq s\leq {\rm
mult}(e)$, as an eigenvalue of $K$, under the perturbation \fer{2}
plus \fer{3}.\footnote{To be more precise, one has to use a
`spectral deformation' $K_\theta$ of the operator $K$ in this
argument \cite{MSB1}, but the deformation does not influence the
physical results.} We have $\varepsilon_e^{(s)}\neq
\varepsilon_e^{(s')}$ unless $s=s'$.
 The $Q_e^{(s)}$ are the (not orthogonal) spectral projections of $K$,
 and $\gamma>0$ satisfies $0\leq\Im\varepsilon_e^{(s)}<2\gamma<T$ (temperature).

We now describe the perturbation expansion  in $\alpha$ of
$\varepsilon_e^{(s)}$ and $Q_e^{(s)}$.  Due to Assumption
\fer{assumption} the eigenspace of $L_\ss$ associated to an
eigenvalue $e$ is obtained as follows. Associated to $e$ are unique
indices $1\leq j_1<j_2<\cdots <j_{N_0(e)}\leq N$ (recall \fer{N0})
satisfying $\sigma_j=\tau_j \Leftrightarrow
j\in\{j_1,\ldots,j_{N_0(e)}\}$ for any $(\usigma,\utau)$ with
$e(\usigma,\utau)=e$. Let
$\urho=(\varrho_1,\ldots,\varrho_{N_0(e)})\in \{-1,+1\}^{N_0(e)}$
and define vectors in $\cx^2$ by
\begin{eqnarray}
\xi_j^{\varrho_j} &=&
\xi_j^\pm=
\left[\begin{array}{c} 1\\\alpha_j^\pm\end{array}\right]\label{xij},\\
\widetilde\xi_j^{\varrho_j} &=&
\widetilde\xi_j^\pm=\frac{1}{1+c_j[(\alpha_j^\pm)^*]^2}
\left[\begin{array}{c}
1\\c_j(\alpha_j^\pm)^*\end{array}\right]\label{wxij},
\end{eqnarray}
according to whether $\varrho_j=\pm 1$. Here, $c_j$ is given in \fer{cj} and
\begin{equation}
\alpha_j^\pm = 1+\i \frac{z^\pm_j-a_j}{b_jc_j}
\end{equation}
with $a_j$, $b_j$, $z_j^\pm$ from \fer{aj}, \fer{bj} and \fer{zj}.

Given $e,\urho$, set
\begin{eqnarray}
\eta_e^{(\urho)} &=& \varphi_{\sigma_1,\tau_1}\otimes\cdots\otimes \xi_{j_1}^
{\varrho_1}\otimes\cdots\otimes \xi_{j_{N_0(e)}}^{\varrho_{N_0(e)}}\otimes\cdots
 \otimes\varphi_{\sigma_N,\tau_N}\label{evect},\\
\widetilde\eta_e^{(\urho)} &=&
\varphi_{\sigma_1,\tau_1}\otimes\cdots\otimes
\widetilde\xi_{j_1}^{\varrho_1}\otimes\cdots\otimes
\widetilde\xi_{j_{N_0(e)}}^ {\varrho_{N_0(e)}}\otimes\cdots
\otimes\varphi_{\sigma_N,\tau_N}, \label{evecttilde}
\end{eqnarray}
where
$\varphi_{\sigma_k,\tau_k}=\varphi_{\sigma_k}\otimes\varphi_{\tau_k}\in\cx^2$,
and where at locations $j_k$, we replace $\varphi_{\sigma_k,\tau_k}$
by $\xi$ (or  $\widetilde\xi$) with the appropriate value of
$\varrho_k$.

Let $h$ be a form factor. We define
\begin{equation}
{\cal G}_h(u)=\int_{S^2} |h(u,\Sigma)|^2\d\Sigma,\qquad \mbox{and} \qquad \gamma_+(h)=\lim_{u\rightarrow 0_+} u{\cal G}_h(u).
\end{equation}
Let $\{h_n\}$ and $\{\alpha_n\}$ be form factors and coupling constants, respectively. For an eigenvalue $e$ as in \fer{4}, set
\begin{eqnarray}
x_e(\{\alpha_n\}, \{h_n\}) &=& -\frac{1}{8} \sum_{\{n: \sigma_n\neq\tau_n\}}
\alpha_n^2\sigma_n \, {\rm P.V.}\int_\rx u^2\frac{\cg_{h_n}(|u|)}{u+\hhbar\omega_n}
 \coth(\beta|u|/2)\d u\quad\\
y_e(\{\alpha_n\}, \{h_n\}) &=& \frac{\pi}{8} \sum_{\{n:\sigma_n\neq\tau_n\}}\alpha^2_n
(\hhbar\omega_n)^2 \cg_{h_n}(\hhbar\omega_n) \coth(\beta\hhbar\omega_n/2),\\
y'_e &=&\frac{\pi}{2\beta}\sum_{\{n: \sigma_n\neq\tau_n\}} \nu^2_n\ \gamma_+(f_n),\\
y''_e&=& \frac{\pi}{8\beta}\gamma_+(f_{\c}) [e_0(e)]^2,\\
e_0(e) &=& \sum_{\{n: \sigma_n\neq\tau_n\}}\varkappa_n\, (\sigma_n-\tau_n).\label{enot}
\end{eqnarray}
Note that the indices over which the sums are taken are the same for any pair of spin configurations
$(\usigma,\utau)$ with $e(\usigma,\utau)=e$. Furthermore, we define
\begin{eqnarray}
X_e&=& x_e(\{\lambda_n\},g_\c) +x_e(\{\mu_n\},\{g_n\}),\\
Y_e&=& y''_e+y'_e+ y_e(\{\lambda_n\},g_\c) +y_e(\{\mu_n\},\{g_n\}).
\end{eqnarray}
Then we have:
\begin{prop}
\label{proposition}
Suppose that the numbers \ $e+\delta_e^{(\urho)}$, where
\begin{equation}
\delta_e^{(\urho)} = X_e +\i Y_e +\sum_{k=1}^{N_0(e)} z_{j_k}^{\varrho_k},
\label{delta}
\end{equation}
are distinct for all $e$ and all $\urho$ (the $z$ are given in
\fer{zj}). Then, for nonzero, small $\alpha$, the eigenvalues of
(the spectrally deformed) $K$ are all simple and have the  expansion
\begin{equation}
\varepsilon_e^{(\urho)} = e +\delta_e^{(\urho)} +O(\alpha^4)
\label{epsilon}
\end{equation}
with corresponding eigenprojection
\begin{equation}
Q_e^{(\urho)} = |\eta_e^{(\urho)}\rangle\langle\widetilde \eta_e^{(\urho)}| +O(\alpha^2).
\end{equation}
\end{prop}
We give a proof of the proposition in Section \ref{lsosect}.
Combining the result of the Proposition with \fer{6} and \fer{7}
gives
\begin{equation}
\av{A_1}_t = \sum_e\sum_{\urho\in\{\pm 1\}^{N_0(e)}} \e^{\i t\varepsilon_e^{(\urho)}}
\scalprod{\psi_{\s_1}\cdots\psi_{\s_N}}{B_1\cdots B_N (|\eta_e^{(\urho)}
\rangle\langle\widetilde \eta_e^{(\urho)}|) A_1\Omega_\ss} +O(\alpha^2),
\label{10}
\end{equation}
with a remainder term uniformly bounded in $t\geq 0$. Since
$\widetilde\eta_e^{(\urho)}$  belongs to the range of the spectral
projection $P(L_\ss=e)$, and since
\begin{equation}
A_1\Omega_\ss = P(L_{\s_2}=\cdots=L_{\s_N}=0) A_1\Omega_\ss,
\end{equation}
only the terms $e\in {\rm
spec}(L_1)=\{-\hhbar\omega_1,0,0,\hhbar\omega_1\}$ in the sum in
\fer{10} contribute.

Let us first consider $e=-\hhbar\omega_1$. We have $N_0(\hhbar\omega_1)=N-1$,
\begin{equation}
\eta_{-\hhbar\omega_1}^{(\urho)} =\varphi_{+-}\otimes\xi_2^{\varrho_2}\otimes\cdots\otimes \xi_N^{\varrho_N} \quad \mbox{and}\quad
\widetilde\eta_{-\hhbar\omega_1}^{(\urho)} = \varphi_{+-}\otimes\widetilde\xi_2^{\varrho_2}\otimes\cdots\otimes \widetilde\xi_N^{\varrho_N}.
\end{equation}
The term with $e=-\hhbar\omega_1$ in \fer{10} equals
\begin{eqnarray}
\lefteqn{
\sum_{\varrho_2,\ldots,\varrho_N\in\{\pm1\}}\e^{\i t[-\hhbar\omega_1 +X_{-\hhbar\omega_1} +\i Y_{-\hhbar\omega_1} +\sum_{j=2}^N z_j^{\varrho_j} +O(\alpha^4)]}}\\
&&\times [\rho_0^{(1)}]_{21}[A_1]_{12} \prod_{j=2}^N\scalprod{\psi_{\s_j}}{B_j\xi_j^{\varrho_j}}\scalprod{\widetilde\xi_j^{\varrho_j}}{\Omega_{\s_j}}\\
&=& \e^{\i t(-\hhbar\omega_1+X_1+\i Y_1 +O(\alpha^4))} {\cal C}_1(N,t)\, [\rho_0^{(1)}]_{21}[A_1]_{12},
\label{001}
\end{eqnarray}
where we set
\begin{equation}
{\cal C}_1(N,t) = \prod_{j=2}^N\left[ \e^{\i tz_j^+} \scalprod{\psi_{\s_j}}{B_j\xi_j^+}\scalprod{\widetilde\xi_j^+}{\Omega_{\s_j}} +
 \e^{\i tz_j^-} \scalprod{\psi_{\s_j}}{B_j\xi_j^-}\scalprod{\widetilde\xi_j^-}{\Omega_{\s_j}}
\right].
\label{c1nt}
\end{equation}
Let us analyze the factors of this product. Using \fer{xij} and \fer{wxij} we have (omitting the index $j$)
\begin{eqnarray}
\lefteqn{
\scalprod{\psi_{\s}}{B\xi^{\varrho}}\scalprod{\widetilde\xi^{\varrho}}{\Omega_{\s}}}\\
&=&
\frac{1}{1+c[\alpha^\varrho]^2}\scalprod{\psi_{\s}}{B(\varphi_{11}+\alpha^\varrho\varphi_{22})}\scalprod{\varphi_{11}+c[\alpha^\varrho]^*\varphi_{22}}{2^{-1/2}(\varphi_{11}+\varphi_{22})}
\nonumber\\
&=&\frac{1+c\alpha^\varrho}{1+c[\alpha^\varrho]^2} \scalprod{\psi_\s}{B \,2^{-1/2}(\varphi_{11}+\alpha^\varrho\varphi_{22})} \nonumber\\
&=&\frac{1+c\alpha^\varrho}{1+c[\alpha^\varrho]^2} \scalprod{\psi_\s}{B \{ |\varphi_1\rangle\langle\varphi_1|\otimes\bbbone +\alpha^\varrho|\varphi_2\rangle\langle\varphi_2|\otimes\bbbone\}\Omega_\s}\\
&=&\frac{1+c\alpha^\varrho}{1+c[\alpha^\varrho]^2} \big( [\rho_0]_{11} +\alpha^\varrho [\rho_0]_{22}\big).
\end{eqnarray}
In the last step, we use that $B$ commutes with $|\varphi_j\rangle\langle\varphi_j|\otimes\bbbone$, and that $B\Omega_\s=\psi_\s$.  Next we note the relation $\alpha^+\alpha^-=-1/c$, which can be derived readily, for instance from the fact that $|\xi^+\rangle\langle\widetilde\xi^+|+|\xi^-\rangle\langle\widetilde\xi^-|=\bbbone$. A short calculation then shows that
$$
\zeta:=\frac{1+c\alpha^+}{1+c(\alpha^+)^2} = 1- \frac{1+c\alpha^-}{1+c(\alpha^-)^2},
$$
so that the factor in the product \fer{c1nt} becomes
$$
{}[\rho_0]_{11}\left\{\e^{\i t z^+}\zeta +\e^{\i tz^-}(1-\zeta)\right\} + (1-[\rho_0]_{11}) \left\{ \e^{\i t z^+}\alpha^+\zeta +\e^{\i tz^-}\alpha^-(1-\zeta)\right\}.
$$
Next, collecting the terms proportional to $[\rho_0]_{11}$ and using
$$
(1-\alpha^-)(1-\zeta) = -(1-\alpha^+)\zeta \mbox{\ \ and\ \ } \alpha^+\zeta = 1-\alpha^-(1-\zeta),
$$
we obtain formula \fer{calcj}.

One can transfer the error term down from the exponent: with $D=O(\alpha^4)$
 we have $\e^{-tY_1 +t D} -\e^{-t Y_1} = \e^{-t Y_1}\sum_{n\geq 1} \frac{(tD)^n}{n!}$ and hence
\begin{equation}
|\e^{-tY_1 +t D} -\e^{-t Y_1}|\leq \e^{-t Y_1} [\e^{t|D|}-1]\leq \e^{-tY_1} t|D| \e^{t|D|}
\end{equation}
(mean value theorem). Now for $|D|\leq C\alpha^4$ and $Y_1\geq c\alpha^2>0$, the r.h.s. can be bounded from above as follows: let $\epsilon>0$, then for $\alpha\leq c\epsilon/C$, an upper bound is
$$
Ct\alpha^4 \e^{-t\alpha^2c(1-\epsilon)} \leq C\alpha^2\e^{-t\alpha^2c(1-2\epsilon)} \sup_{x\geq 0} x \e^{-xc\epsilon}= \frac{C\alpha^2}{e c\epsilon}\e^{-t\alpha^2c(1-2\epsilon)}.
$$
This gives that if $Y_1\geq c\alpha^2>0$, then for all $\epsilon>0$ and $\alpha$ small enough,
\begin{equation}
\e^{-t Y_1+O(\alpha^4)} =\e^{-t Y_1} +O(\alpha^2\e^{-tc\alpha^2(1-\epsilon)}).
\end{equation}
The remainder depends on $\epsilon$. Taking $\epsilon=1/2$ and $\alpha<c/(2C)$, we get
\begin{equation}
\e^{-t [Y_1+O(\alpha^4)]} =\e^{-t Y_1} +O(\alpha^2)
\end{equation}
(uniformly in $t\geq 0$). This gives the contribution \fer{f1}.

Similarly to \fer{001}, one shows that the term in \fer{10} with $e=\hhbar\omega_1$ equals \fer{f2}. To derive this, one checks that under the change $e\mapsto -e$, the exponent in \fer{001} undergoes a complex conjugation and a sign change, and ${\cal C}_1$ turns into its complex conjugate.

Next we conisder $e=0$ in \fer{10}. We have $N_0(0)=N$ and obtain two contributions: one associated with $\varepsilon_0=0$, $\eta_0=\Omega_\ss$ (see \fer{kernel}), $\widetilde\eta_0=\widetilde\xi_1^+\otimes\cdots \otimes\widetilde\xi_N^+$ and another contribution with $\varepsilon_0^{(-1,1,1,\ldots)}$ and
\begin{equation}
\eta_0^{(-1,1,1,\ldots)}=\xi_1^-\otimes\xi_2^+\otimes\cdots\otimes\xi_N^+\quad \mbox{and}\quad \widetilde\eta_0^{(-1,1,1,\ldots)}=\widetilde\xi_1^-\otimes\widetilde\xi_2^+\otimes\cdots\otimes\widetilde\xi_N^+.
\end{equation}
It is easily seen that these two contributions are \fer{f0} and \fer{f00}. This completes the proof of Theorem \ref{thm1} \hfill $\blacksquare$

\subsection{Level shift operators and proof of Proposition \ref{proposition}}
\label{lsosect}

The total Liouville operator has the form
\begin{equation}
K = L_0+ W,
\end{equation}
where $L_0=L_\ss+L_\rr$ is the free (non interacting) Liouville operator, and $W$ contains all interactions and is of $O(\alpha^2)$ (for more detail, see Appendix B of \cite{MSB1}, Resonance theory of decoherence and thermalization). To every eigenvalue $e$ of $L_0$ we associate the level shift operator
\begin{equation}
\Lambda_e = -P_e W \overline P_e \,(L_0-e+\i 0_+)^{-1}\overline P_e W P_e,
\label{deflso}
\end{equation}
where $P_e$ is the spectral projection of $L_0$ onto the eigenvalue $e$. The eigenvalues of $\Lambda_e$ are the second order (in $\alpha$) corrections to the eigenvalues of $K$ under the analytic (in $\alpha$) perturbation $W$ of $L_0$, see also Section 5 of \cite{MSB1}, Resonance theory of decoherence and thermalization. Moreover, if $\varepsilon_e^{(\urho)}=e+\delta_e^{(\urho)}+O(\alpha^4)$ are the eigenvalues bifurcating out of $e$ for $\alpha\neq 0$ then the corresponding eigenprojections of $K$, to lowest order in $\alpha$, are given by $|\eta_e^{(\urho)}\rangle\langle\widetilde\eta_e^{(\urho)}|$, where
\begin{eqnarray}
\Lambda_e \eta_e^{(\urho)} &=& \delta_e^{(\urho)}\eta_e^{(\urho)}\\
{}[\Lambda_e]^*\widetilde \eta_e^{(\urho)} &=& [\delta_e^{(\urho)}]^*\widetilde\eta_e^{(\urho)}
\end{eqnarray}
and $\scalprod{\eta_e^{(\urho)}}{\widetilde \eta_e^{(\urho)}}=1$. For more information on these facts, we refer to Section 6 of \cite{MSB1}, Resonance theory of decoherence and thermalization. We are assuming here that all energies $\delta_e^{(\urho)}$ are different, so that the corresponding eigenspaces are one-dimensional. This is generically true in applications, but it is not necessary for our strategy to work, see e.g. Appendix A of \cite{MSB1}, Dynamics of collective decoherence and thermalization.

Below, we give the explicit form of the level shift operators associated with all eigenvalues $e$ (given by\fer{4}). Each level shift operators splits into a sum
\begin{equation}
\Lambda_e = \Lambda_e^{\rm coll} +\Lambda_e^{\rm loc}
\label{totallso}
\end{equation}
of two operators associated with the local and the collective interactions. We find the spectrum and eigenvalues of the level shift operators. In view of the explanations given at the beginning of this section, this gives a proof of Proposition \ref{proposition}.

\subsubsection{Collective level shift operator}

Let $e$ be an eigenvalue \fer{4} and define
\begin{equation}
r_n= \frac{1}{4} \varkappa_ne_0(e)\, {\rm P.V.} \int_{\rx^3}\frac{|f_{\c}(k)|^2}{|k|}\d^3k
\end{equation}

\begin{prop}[Collective LSO]
\label{colllsoprop}
The collective level shift operator associated to $e$ is given by
\begin{equation}
\Lambda_e^{\rm coll} = x_e(\{\lambda_n\}, g_\c)+\i[y''_e+y_e(\{\lambda_n\},g_\c)] + \sum_{\{n: \sigma_n=\tau_n\}} M_{\rm coll}^n,
\end{equation}
where $M_{\rm coll}^n$ acts on ${\rm span}\{\varphi_{++},\varphi_{--}\}$ (doubled Hilbert space of $n$-th spin) as
\begin{equation}
M_{\rm coll}^n = \i\frac{\pi}{4} \lambda_n^2 (\hhbar\omega_n)^2\frac{\cg_{h_n}(\hhbar\omega_n)}{\e^{\beta\hhbar\omega_n}-1}
 \left[
\begin{array}{cc}
1 & -1\\
-\e^{\beta\hhbar\omega_n} & \e^{\beta\hhbar\omega_n}
\end{array}
\right]
-r_n
\left[
\begin{array}{cc}
1 & 0\\
0 & -1
\end{array}
\right].
\end{equation}
\end{prop}
A proof of this proposition is obtained along the lines of Proposition 3.7 of \cite{MSB1} (Dynamics of collective decoherence and thermalization).

\subsubsection{Local level shift operator}

\begin{prop}[Local LSO]
\label{loclsoprop}
The local level shift operator associated to $e$ is given by
\begin{equation}
\Lambda_e^{\rm loc} = x_e(\{\mu_n\},\{g_n\}) +\i[ y'_e + y_e(\{\mu_n\}, \{g_n\})] + \sum_{\{n: \sigma_n=\tau_n\}} M_{\rm loc}^n,
\end{equation}
where $M_{\rm loc}^n$ acts on ${\rm span}\{\varphi_{++},\varphi_{--}\}$ (doubled Hilbert space of $n$-th spin) as
\begin{equation}
M_{\rm loc}^n =  \i\frac{\pi}{4} \mu_n^2 (\hhbar\omega_n)^2\frac{\cg_{g_n}(\hhbar\omega_n)}{\e^{\beta\hhbar\omega_n}-1}
\left[
\begin{array}{cc}
1& -1\\
-\e^{\beta\hhbar\omega_n} & \e^{\beta\hhbar\omega_n}
\end{array}
\right].
\end{equation}
\end{prop}
A proof of this proposition is obtained along the lines of Proposition 5.1 of \cite{MSB1} (Resonance theory of decoherence and thermalization).

\bigskip

{\bf Remark.\ } The contributions to local and collective level shift operators coming from the energy-exchange interactions are the same.

\subsubsection{Proof of Proposition \ref{proposition}}

The explicit forms of $\Lambda_e^{\rm coll}$ and $\Lambda_e^{\rm loc}$ given in Propositions \ref{colllsoprop} and \ref{loclsoprop}, and relation \fer{totallso} yield
\begin{equation}
\Lambda_e = X_e+\i Y_e +\sum_{\{n:\sigma_n=\tau_n\}}\left\{ \i b_n
\left[
\begin{array}{cc}
c_n & -c_n\\
-1 & 1
\end{array}
\right] - r_n
\left[
\begin{array}{cc}
1 & 0\\
0 & -1
\end{array}
\right]
\right\}.
\end{equation}
The results \fer{epsilon}, \fer{delta} and \fer{evect} for the eigenvalues and eigenvectors follow, and similarly for its adjoint.

Note that for $e=0$ we have $e_0(e)=0$, and the level shift operator becomes
\begin{equation}
\Lambda_0 =\i \sum_{n=1}^N b_n
\left[
\begin{array}{cc}
 c_n & -c_n\\
-1 & 1
\end{array}
\right].
\end{equation}
It follows that $z_n^+=0$ and $z_n^-=\i b_n(c_n+1)$, with
\begin{equation}
\xi_n^+ = \frac{1}{\sqrt 2}\left[
\begin{array}{c}
1\\
1
\end{array}
\right], \ \
\xi_n^- = \left[
\begin{array}{c}
c_n\\
-1
\end{array}
\right], \ \
\widetilde\xi_n^+ = \frac{\sqrt 2}{1+c_n}\left[
\begin{array}{c}
1\\
c_n
\end{array}
\right], \ \
\widetilde\xi_n^- = \frac{1}{1+c_n}\left[
\begin{array}{c}
1\\
-1
\end{array}
\right].
\end{equation}

\section{Validity of perturbation expansion}
\label{sec5}

The Heisenberg equations of motion corresponding to \fer{1}-\fer{3} are
\begin{eqnarray}
\dot S^z_n &=&-\i [S_n^z,H] = \lambda_\c S_n^y\otimes \phi_\c(g_\c) +\mu_n S^y_n\otimes\phi_n(g_n)\\
\dot S^x_n &=&\omega_n S_n^y -\varkappa_n S_n^y\otimes \phi_\c(f_\c) -\nu_n S^y_n\otimes\phi_n(f_n)\label{ex}\\
\dot S^y_n &=&-\omega_n S_n^x -\lambda_n S_n^z\otimes \phi_\c(g_\c) +\varkappa_n S^x_n\otimes\phi_\c(f_\c) -\nu_n S^z_n\otimes\phi_n(g_n) \\
&& +\nu_n S_n^x\otimes\phi_n(f_n).
\end{eqnarray}
For the local and collective annihilation operators $a_n(k)$ and $a_\c(k)$ we have
\begin{eqnarray}
\i \dot a_n(k) &=& [a_n(k),H] = |k| a_n(k) +\frac{1}{\sqrt 2}\mu_n g_n(k) S^x_n +\frac{1}{\sqrt 2}\nu_n f_n(k) S^z_n \\
\i \dot a_\c(k) &=&|k| a_\c(k) +\frac{1}{\sqrt 2}\lambda_n g_\c(k) \sum_{n=1}^NS^x_n +\frac{1}{\sqrt 2}\varkappa_n f_\c(k) \sum_{n=1}^NS^z_n.
\end{eqnarray}
The latter two equations can be integrated,
\begin{eqnarray}
a_n(k,t) &=& \e^{-\i|k|t} a_n(k,0)\nonumber\\
& &  - \frac{1}{\sqrt 2}\int_0^t \e^{-\i |k|(t-\tau)} \left\{\mu_n g_n(k) S^x_n(\tau) +\nu_n f_n(k) S^z_n(\tau)\right\}\d\tau\label{e1}\\
a_\c(k,t) &=& \e^{-\i|k|t} a_\c(k,0)\nonumber\\
& &  -\frac{1}{\sqrt 2}\int_0^t \e^{-\i |k|(t-\tau)} \left\{\lambda_n g_\c(k) \sum_{n=1}^N S^x_n(\tau) +\varkappa_n f_\c(k) \sum_{n=1}^NS^z_n(\tau)\right\}\d\tau.\ \ \ \label{e2}
\end{eqnarray}
Remembering that $\phi(h)=\frac{1}{\sqrt 2}\int_{\rx^3} \{ h(k) a^*(k)+\overline{h}(k) a(k)\}\d^3k$ we insert \fer{e1}, \fer{e2} into \fer{ex} to obtain
\begin{eqnarray}
\dot S^x_n(t) &=& \omega_n S_n^y(t) -\varkappa_n S_n^y(t)\phi(\e^{\i |k|t}f_\c)\nonumber\\
&&\hspace*{-1.5cm}+\frac 12 \varkappa_nS^y_n(t) \int_{\rx^3}\int_0^t\e^{-\i|k|(t-\tau)}\overline{f_\c}(k)\left\{\lambda_n g_\c(k) \sum_{n=1}^N S^x_n(\tau) +\varkappa_n f_\c(k) \sum_{n=1}^NS^z_n(\tau)\right\}\d\tau\d^3k\nonumber\\
&&+{\rm h.c.}\nonumber\\
&&\hspace*{-1.5cm}+\frac 12 \nu_nS^y_n(t) \int_{\rx^3}\int_0^t\e^{-\i|k|(t-\tau)}\overline{f_n}(k)\left\{\mu_n g_n(k) S^x_n(\tau) +\nu_n f_n(k) S^z_n(\tau)\right\}\d\tau\d^3k\nonumber\\
&&+{\rm h.c.}\label{e5}
\end{eqnarray}
Let us denote by $h^2(k)$ any of the products of functions of $k$ occuring in the above integrals (e.g. $h^2(k)=\overline{f_\c}(k)g_\c(k)$ etc). Let us analyze the $k$-integrals in the last expression for $\dot S^x_n$. The product of form factors behaves like $h^2(k)=|k|^{p_1+p_2}\e^{-|k|/k_0}$, where $p_j=-1/2+n_j$, $n_j=0,1,2,\ldots$ (see Assumption (C)). So
$$
\int_{\rx^3}\e^{\i|k|\tau}h^2(k)\d^3k \sim\int_0^\infty r^{1+n_1+n_2} \e^{\i r\tau} \e^{-r/k_0}\d r=\partial_\tau^{1+n_1+n_2}\frac{(-\i)^{1+n_1+n_2}}{1/r_0+\i\tau},
$$
which decays at least as $\frac{1}{1+\tau^2}$ (worst case $n_1=n_2=0$). Together with the boundedness $|S^{x,y,z}_n(\tau)|\leq 1/2$ this implies that the integrals over $\tau$ and $k$ in \fer{e5} are bounded homogeneously in $t\geq 0$, leading to
$$
\dot S^x_n(t) = \omega_n S_n^y(t) -\varkappa_nS_n^y(t)\phi(\e^{\i|k|t}f_\c) +\varkappa_nO(\lambda_nN +\varkappa_nN) +\nu_n O(\mu_n+\nu_n).
$$
For the validity of perturbation theory homogeneously in $t\geq 0$, we should impose $\varkappa^2_nN, \varkappa_n\lambda_nN<\!\!<\omega_n$ and $\nu_n^2, \nu_n\mu_n<\!\!<\omega_n$. Denoting by $\alpha_\c$ and $\alpha_\ell$ the size of collective and local coupling parameters, we thus need $\alpha_\c^2N<\!\!<\omega$, $\alpha_\ell<\!\!<\omega$, where $\omega$ is the (typical) Bohr frequency of the single spin.

\bigskip
\bigskip

\noindent
{\Large\bf Acknowledgements}

\bigskip
\noindent We would like to thank M.A. Espy and P.L. Volegov for
useful discussions. This work was carried out under the auspices of
the National Nuclear Security Administration of the U.S. Department
of Energy at Los Alamos National Laboratory under Contract No.
DE-AC52-06NA25396. This research by G.P. Berman was supported by the
LDRD Program at LANL and  by the Office of the Director of National
Intelligence (ODNI), and Intelligence Advanced Research Projects
Activity (IARPA). All statements of fact, opinion or conclusions
contained herein are those of the authors and should not be
construed as representing the official views or policies of IARPA,
the ODNI, or the U.S.M.\  M. Merkli acknowledges the support of
NSERC under Discovery Grant 205247, and of the Quantum Institute
through the CNLS at LANL.

\end{document}